# Multi Relational Data Mining Approaches: A Data Mining Technique

Neelamadhab Padhy
Asst.Professor, GIET, Gunupur.
Research Scholar, CMJ University, Shilong (Meghalaya),

Rasmita Panigrahi
Lecturer, G.I.E.T, Gunupur
Department of IT.
G.I, E.T, Orissa
India

## ABSTRACT

The multi relational data mining approach has developed as an alternative way for handling the structured data such that RDBMS. This will provides the mining in multiple tables directly. In MRDM the patterns are available in multiple tables (relations) from a relational database. As the data are available over the many tables which will affect the many problems in the practice of the data mining. To deal with this problem, one either constructs a single table by Propositionalisation, or uses a Multi-Relational Data Mining algorithm.MRDM approaches have been successfully applied in the area of bioinformatics. Three popular pattern finding techniques classification, clustering and association are frequently used in MRDM. Multi relational approach has developed as an alternative for analyzing the structured data such as relational database. MRDM allowing applying directly in the data mining in multiple tables. To avoid the expensive joining operations and semantic losses we used the MRDM technique. This paper focuses some of the application areas of MRDM and feature directions as well as the comparison of ILP, GM, SSDM and MRDM.

## 1. INTRODUCTION

The main objective of the data mining techniques is to extract regularities from a large amount of data. Data mining has a variety of fields which provides the different tools and the techniques for handling the large database. Through this technique we will obtain the new, valuable non-trivial and existing information [1] [34]. As we know that the relational database are the best approach to handle the structured data but now a days the data are not only available in relational form but also available in multiple form. There are different types of data mining task can be performed in the data base .These tasks are forming a group which is called as cluster or to identify the individual piece of data called as outlier which does not fit with other data. These databases are connected but using the entity-relationship links in the database design. [2]."To extract the regularities directly from a database without having Structure data "which leads to MRDM? The term MRDM means Multi Relational Data Mining. Multi-relational Data Mining (MRDM) aims to discover knowledge directly from relational data. There have been many approaches for Classification, such as neural networks and support vector machines. If we gather all the tables in centrally then which is made up of attributes that summarizes or aggregate the information found in other tables. This technique is obviously a disadvantage because, many attributes and data repetitions occurred [3]. The need of MRDM is increasing the amount of complex data generated which has stored in the form of relational form. **Cios and Moore** pointed out some feature that makes medical data mining unique [4]. Our study is how to convert the relational structures into Multi relational form. **Knobbe et.al [5]** which proposed a generalized framework for MRDM which exploits SQL to gather the information needed for decision tree from Multi relational data .Data mining refers the "extracting" or "mining" knowledge from huge amount of the data .Thus we called the data mining as the "KNOWLEDGE MINING". So finally the data mining in short we called as KDD (Knowledge mining from data). [6][10]. In the real world ,most of the data are stored in multiple relations .the MRDM aims to discovering interesting knowledge directly from multiple tables without joining the data of multiple tables into a single table explicitly.[11][12][13]. In Multi relational data mining where the relations are available in the form of parent and child relationship like primary and foreign key combinations .if we adopt the traditional algorithms then these algorithms are not efficient due to the difficulty of allocating enough memory for all the data structure used for the algorithm to represent the database. The above problem is called the Scalability problem in the MRDM. **Valencio et.al**, *Human-centric computing and Information Science 2012 2:4* was proposed the first time to avoid the scalability problem which occurred in the MRDM. To avoid the above problem they proposed the MR-Radix algorithm.MR-Radix is a multi relational data mining algorithm, which has a data structure called the Radix tree **[7].** Radix tree is one of the data structures which compress the database in the memory. The MR-Radix algorithm provides the better efficiency when we compared with the traditional algorithms. Suppose the data is scattered over many tables causes many Problems in the practice of data mining. To avoid the above problem *Krogel, and Knobbe* defined that how the aggregate functions are used in the Multi Relational data mining **[8] [9].** In addition to that when we deal with the logic–based MRDM, which is called as Inductive Logic Programming **[13] [14] [15] [16]**

This paper is a summary of previous work **[2, 4, 5, 7, 11, 10, and 12].**

This paper emphasizes 10 sections. Section 1 describes the introductory portion where we have focused the works already done and Section 2 indicates that the main motivation of the MRDM and Section 3 provides its framework of the MRDM.

Section 4 describes the different techniques of data mining system; section 5 provides the structured data mining and its classification. In Section 6 we have given the analysis of MRDM parameters as well as some approaches and finally section 7 and 8 we have included the feature directions of some applications of Mufti Relational data mining. A novel work has been done in the section 9 which is describing the details survey along with a tabular representation. Section 10 provides the comparisons ILP, GM, SSD and MRDM along with the 6 properties i.e. existing method and Proposed Method.

## 2 MRDM MOTIVATION

Now a days many algorithm which are completely based on the attribute – value setting but these algorithms restricts their use to data sets consisting of single table. The motivation of MRDM is how the structured data are stored in the relational database. MRDM can perform the linkage Knowledge discovery in multi relational environments, Multi relational Rules, Multi relational clustering, Multi Relational Clustering, Multi Relational Linkage analysis.

### 2.1. MRDM APPROACHES

There are so many approaches are supported by the Multi Relational Data Mining these are as below:

- Inductive Logic Programming (ILP)
- Multi-relational Clustering
- Probabilistic Relational Models

## 3 MRDM FRAME WORK

MRDM is a multi-disciplinary field which dealing with the Knowledge discovery from relational database which consisting of number of relations. It is frameworks which deals with gathering the data about the data (metadata) from a database and choose the best approach to get the optimal results. This framework is useful to the researchers .Before implement any algorithm, let this framework says that Data Understanding and Data Preparation is highly necessary. MRDM aims to integrate the results from existing fields like ILP, KDD, Statistics, Machine learning. Data understanding means gathering the metadata from the database and which describe the best approach of the analysis. Data Preparation means transformation of the database into MRDM formats where we select the algorithms.

### 3.1. Data Understanding Tasks:

- Selects the tables
- Determine the background
- Determine the individual
- Determine the target

### 3.2 Data Preparation Tasks:

- To construct the feature
- Discretise numeric data
- Denormalise
- Aggregate and propositionalise

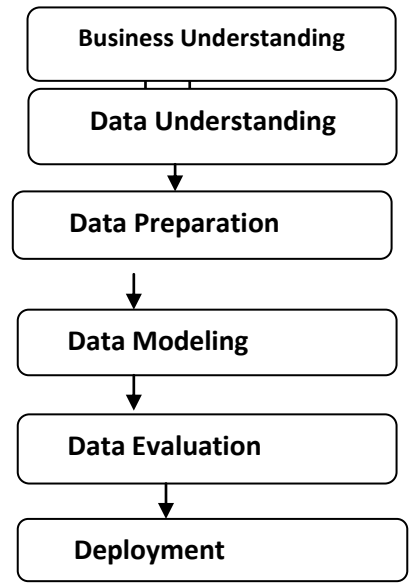

**Figure - 1 (MRDM Framework Architecture**

Multi-relational data mining framework is based on the Search for interesting patterns in the relational database, Where multi-relational patterns can be viewed as" pieces of Substructure encountered in the structure of the objects of interest" [Knobbe *et al.*, 1999a].

## 4 DATA MINING TECHINQUES

Data Mining Technique can be broadly divided into two categories .These are as below

- Unique Table Approach Mining
- Structured Data Mining

### 4.1. Unique Table Approach Mining.

This unique table approach mining is popularly known as propositional data mining. The important task of the data mining is known as Propositional Data Mining. The aim assumption is that each individual is represented by a fixed set of characteristics which is known as attributes. Again the individual can be thought of as a collection of attribute –value pairs, which are represented as a vector format [17]

In this approach the central database of individuals becomes a table: which consists of rows(or tuples) correspond to individuals and columns (attribute).If we want to establish the relation among the different tables (which may in the form of Primary and Foreign key relationships) then we have encounter the two problems .There are two ways to handle the problem

**First Problem**: Compute the join but this leads to data redundancy, Missing values. A single instance in the original database is mapped onto multiple instances in the new table, which is the problematic.

**Second Problem**:

The second way is aggregate the information in different tuples representing the same individual into one tuples after computing the join.

Which may be solved the problem but causes the loss of information.

## 4.2. Structured Data Mining

The term structured data mining means to handle the complex data. That is this kind of data that cannot store in the sensible in one table or doesn't have a single table representation in which table rows or columns are not related to each other. Now a day's everyone using multirelational databases i.e. databases of multiple tables. Some problems involving molecules, proteins, phylogenetic trees, social networks and web server access logs cannot be tackled if rows in a single table are not explicitly linked to each other. To deal with structured data all kinds of problems have to be tackled that are almost trivially solved in attribute-value mining algorithms. In present day's scenario, the amount of data is available hugely so human must take help from automatic computerized methods for extracting the required information. These types of data must be represented in the form of graph data structure which will represent the nodes and their attributes, relationships with the other entities.

## 5 STRUCTURED DATA MINING TYPES

There are 4 categories are available these are as below:

### 1.1. Graph Mining

Graph mining is the techniques which will extract the required information from data represented in the form of graph structured form. A graph can be defined as the equation $G=\{V,E\}$, Where $V =\{v_1,v_2,v_3,\ldots\ldots\ldots\ldots\ldots v_n\}$ is an ordered set of vertices in the graph and $=\{e_1,e_2,e_3,\ldots\ldots e_n\}$ is the set of pair of edges. The term graph mining which can refer to discover the graph patterns. **Chakrabarti & Faloutsos** [18], [42] has defined the two scenarios: First: It is the typical of web domains and the second one is: database of chemical compounds. The main objective of the Graph Mining is the concept of *frequent graph pattern*. Graphs becomes increasingly important in modeling complicated structures, such as circuits, images, chemical compounds, protein structures, biological networks, social networks, the web, workflows and XML documents. Many graphs search algorithms have been developed in chemical informatics, computer vision, video indexing and text retrieval. With the increasing demand on the analysis of large amounts of structured data, graph mining has become an active and important theme inn data mining. Graph mining is used to mine frequent graph patterns and perform characterization, discrimination, classification **[18]** etc.

### 5.2 Inductive Logic Programming (ILP)

This ILP paradigm says that how the logic program will convert the patterns. The logic program induced from a database of logical facts which is called as ILP.ILP follows the top-down approach. The advantage of ILP is too expressive and powerful rules are understandable the disadvantage of ILP is Inefficient for the database with the complex schema as well as not properly handled the continuous attributes.

The database consists of a collection of facts in first-order logic**)** [19], [20]. Each fact represents a part, and individuals can be reconstructed by piecing together these facts. First-order logic (often Prolog) can be used to select subgroups. [21]

### 5.3 Semi-Structured Data Mining

Now a days the data mining is concerned with the discovery of patterns as well as the relations .Almost all the data mining algorithms handle the data with the fixed form but the schema is defined with advance When the data is on the web then its structure is regular form .We normally called as semi-structured data mining .To handle such type of data we use the XML language .XML not only handles the tabular data but also the arbitrary trees. Semi-structured data is naturally modeled in terms of graphs contain labels that give semantics to the underlying structure. The database consists of XML documents, which describe objects in a mixture of structural and free-text information. [22]

### 5.4 Multi-Relational Data Mining (MRDM)

The database consists of a collection of tables (a relational database). Records in each table represent parts, and individuals can be reconstructed by joining over the foreign key relations between the tables. [23][24]**[25][26]**. the database consists of a collection of tables (a relational database). Records in each table Represent parts, and individuals can be reconstructed by joining over the foreign key relations between the tables. Subgroups can be defined by means of SQL or a graphical query language. [27]

### 5.4.1. Selection Graph:

When multi relational patterns are expressed in terms of graphical language then we called as the Selection Graph. Selection graph model is used in SQL (Database Language) to directly deal with tables. There are so many algorithms was developed which aims to convert the selection graph model into SQL query form. [28]

**Anna Atramentov et.al** also focused the graphical representation of selection graph. [29]

### 5.4.2. TupleID Propagation:

It is a method of transferring information among different relations by virtually joins them. This method exhibit to search in the relational database and which is observed that less costly than physical joins in both time and space. It is the technique for performing virtual joins among relations which less expansive then physical joins. When we want to search a good predicate then we will use propagate Tuple Ids between any two relations which provides the less computations and storage cost compared with creating join requirements [11]

During this study we found there are two cases that such propagates could be counterproductive.

**Drawbacks:**

 Propagates via large fan-outs

> Propagate via long weak links

During this study we found the two most challenges these are as below:

**Challenge-1:**

> Finding the Useful Links :

*Solution for Problem-I*:

It is necessary that to find the techniques that can estimate the usefulness of links across tables and then use the most useful links to achieve the better data mining task.

**Challenge-II:**

- Transferring information efficiently :

*Solution for Problem-II*:

We must develop the strategies with as low inter database communication cost as possible

### 5.4.3 Multi view Learning

The multi-view learning problem with n views can be seen as n inter-dependent relations and are thus applicable to multi-relational learning. This is the basic scenario in multi relational learning problem The Multi View Classification (MVC) approach employs the multi-view learning framework to operate directly on multi-relational databases with conventional data mining methods. The approach works as follows

- Information Propagation Stage:
- Aggregation Stage:
- Multiple Views Construction Stage:
- View Validation Stage:
- View Combination Stage:

## 2. ANALYSIS OFMRDM APPROACHES

The below available comparative analysis shows (**Figure-I)** some strong points of Multi View Learning (MVL) compare to other approaches. We have compared with the three most important factors.

**Factor –I:**

The relational database is able to keep its compact representation and normalized structure.

**Factor-II:**

The second is that it uses a framework that is able to directly incorporate any traditional single-table data mining algorithm.

**Factor-III:**

The multi-view learning framework is highly efficient for

Mining relational databases in term of running time

## 7. RELATED WORK AND APPLICATION

### 7.1 Multirelational data mining in medical database

The main objective is to extract probabilistic tree patterns from a database using Grammatical Inference techniques. Again the objective is to see these patterns in multi-relational data mining tasks and how they can model structured data in relational database. This method is based on a representation of the database by a set of trees and the inductive phase consists in first learning a SMTA and generalizing this model relatively to a parameter. This paper presents the application of a method for mining data in a multi-relational database that contains some information about patient's strucked down by chronic hepatitis. We propose to use a representation of the database by trees in order to extract these patterns. Trees provide a natural way to represent structured information taking into account the statistical distribution of the data [30].

### 7.2 A Model for MRDM on Demand Forecasting:

In this paper, two models are proposed, which is pure classification (PC) and Hybrid Clustering (HCC) model.HCC model is designed to be generic solutions for multi-relational data mining. The main aim of hybrid-clustering classification is to divide the expensive classification task into simpler task by creating well-defined clusters in the training data set. Experimental result shows that the Hybrid Clustering classification provides better accuracy as well as Scalability than the Pure Classification. The two algorithm are used these are K-Mean Mode and K-Nearest neighbor classification to get as per the demands of the customer. [31]

### 7.3 A multi-relational Decision Tree Learning Algorithm –Implementation and Experiments

MRDM is used in the multi-relational decision tree learning algorithm(MRDTL-2)[32] which was proposed by Knobbe et.al.[29].He describes some simple technique for speeding up the calculation of sufficient statics for decision trees and related hypothesis classes form multi-relational data. He proposed how to handle the missing values.

During this review and study we found the there are two limitations in MRDTL-2 these are as below.

- Slow running time
- Inability to handle missing attribute values

**Table-1 Comparison and analysis for MRDM Approaches**

| Attributes | Approach Based On Relational Databases | | |
| --- | --- | --- | --- |
| | Graph Mining(SG) | Tuples ID propagation | MVC(Multi View Learning) |
| Scalability | Low | Low | High |
| Expressiveness | High | High | Low |
| Time to Learn | Less | More | Less |
| Numerical Attributes | Yes | Yes | Yes |
| Normalized Structure | No | No | Yes |
| Direct mining on real world Database | No | No | Yes |
| Incorporation of Traditional Single Table Algorithm | No | No | Yes |
| Integration of Advance Techniques | No | No | Yes |
| Learning Using Heterogeneous Classifier | No | No | Yes |
| Learning on Heterogeneous Data | No | No | Yes |
| Support of Incremental Design | Low | Low | High |

## 7.4 Neural Networks in Multi-Relational Data Mining

Multi relational data mining can be used in the field of neural network system but Multi relational data mining frame work is based on the search for interesting patterns in the relational database .The combination of data mining method and neural network model can greatly improve the efficiency of data mining methods, and it has been widely used. **M. RaviSankar** and **P. PremChand** has defined that the neural network is used in the MRDM .It is easy to discover a huge number of patterns in a database where most of these patterns are actually obvious, redundant, and useless or uninteresting to the user. During this study we found that multi-relational data mining and the application of soft computing methodologies which involves the neural networks, genetic algorithms, fuzzy sets, and rough sets. [43]

## 7.5 Biological Applications of Multi-relational data mining

We found some of the lessons and challenges these are as below:

One interesting lesson that biological applications such as pharmacophore discovery have provided is how naturally multi-relational data mining address the multiple instances problem. this has been noted earlier [33] .Recently **,Perlich and Provosts** have provided a useful hierarchy of data mining tasks, based on representation, where multiple instances tasks from a class subsumed by multi-relational task.[29].

- ➤ The multiple distinct sources of information can provide improved accuracy.
- ➤ A biologist can use the decision – tree learner algorithm or a linear regression algorithm.

## 7.6 Knowledge discovery from relational database for inductive logic programming (ILP):

In this paper they have described an approach which uses UML as the common specification language for a large range of ILP engines. Having such a common language will enable a wide range of users, including non-experts, to model problems and apply different engines without any extra effort. The process involves transformation of UML into a language called CDBL

## 8. FEATURE DIRECTION

To get better accuracy result the researcher may use the demand forecasting technique i.e. linear regression and non linear regression technique, support vector machine (SVM).

Apart from w**e** proposed the two techniques which are used in the Neural Network these are

- ➤ Effective Combination of Neural Network and Data Mining Technology.
- ➤ Effective Combination of Knowledge Processing and Neural Computation.
- ➤ Slow Running time:

  During the study of MRDM, it has been stated that MRDM produce the slow running time while upload the data set during the experiments.

- ➤ Inability to handle missing attribute values

  This is the research challenges in the MRDM. It researcher can enhance this work.

- ➤ *The researcher they can use the MR-RADIX algorithm* which is based on Patricia mine algorithm which can extract multi-relational association rules from large relational database instead of traditional algorithm. This algorithm produces the better performance and reduces the amount of used memory space by up to 75% when compared with the traditional Multirelational data mining algorithm [7].

- ➤ When the patterns are represented in the logic form then we called as ILP systems. It is an important subset of first order logic. The first order logic clearly corresponds to concepts of relational databases. LBRC can express more complex patterns. But it needs to transform relational data into logic programs in preprocessing stage, which determines the relatively weak relation in database. This conversion leads to lot of inconvenience in real application.

- *The MRDM approaches may be used in Credit risk evaluation,* where financial data are available .It is referred to as the risk of loss when a debtor does not fulfill its debt contact and is of natural interest to practioner in banks as well as to regulators [40].
- *The researcher may use the MRDM approaches in the field of Heart disease prediction.* There are so many technique are used but none of them used the MRDM.[41]The detection of heart diseases from various factors or a symptoms is a multi-layered issue which is not free from false presumptions often accompanied by unpredictable effects It would better if the researcher will use the demand forecasting technique to know the heart disease by using the K-Mean and k-medoid clustering algorithm .As well as the researchers they can use the two model Hybrid clustering and Pure classification model to forecasting the heart disease[31].

## 9. TABULAR REPRESENTATION OF MRDM SURVEY.

**(Table-2 Tabular Representation of MRDM Survey)**

| Topic | Aims | framework/ Architecture/ Model | Technique/ Approach Algorithm | Schema Used | Extended logical database | Case Study | Tool/ DATABASE |
|---|---|---|---|---|---|---|---|
| 01 | 02 | 03 | 04 | 05 | 06 | 07 | 08 |
| et.al.Multi relational classification | Discovering the useful pattern across multiple inter-connected tables (relations) in a relational data. | | Multi view Learning | Star | Yes | - | WEKA |
| et.al.Multi Relational data mining in the field of Medical" | To extract the probabilistic tree patterns from a database | | Grammatical inference technique by using ILP (Stochastic Tree automata ) | Decision Tree structure. | Yes | Medical database are used | Language Bias –tree structure. |
| et.al Knowledge Discovery from relational database for Inductive Logic Program(ILP) | Transformation of UML into a language which is aims to represent the CDBL. | | ILP | UML | yes | - | WARMR |
| et.al A model for Demand forecasting by using MRDM | To prediction of the demands by using the facts. | In this paper used the two models : Pc: Pure classification HCC: Hybrid Clustering | The two algorithms are used : K-Mean Mode K-Nearest Neighbor Classifier | - | yes | | Oracle-9i set up sqlnet.ora and tnsnames.ora with Java tech. |
| Et.al. Knobbe – Multi relational Decision Tree Learning Algorithm. | To speed up of the calculation of statistic from Decision tree. Aims to deal with the Missing values. | MRDM (Multi Relational Data Mining) Framework | Decision Tree Construction  Translating of selection Graph into the SQL Query | snowflake | Yes | - | ILP  Data set :- Mutagenesis KDD CUP(2001) PKDD 2001 |
| et .al. Wenxiang Dou "Distributed Multi Relational Data Mining on GA"(Genetic Algorithm) | To mine implements rules in Multiple tables and combine the method with the GA to enhance the mining efficiency | System architecture of distributed mining for the Multi relational Data mining | Genetic Algorithm , Apriori Algorithm | Star Schema | North wind Data base | - | SQL Server 2000 and they have used Fuzzy Logic |
| et.Valvalencio.MR-Radix: A Multi Relational Data Mining Algorithm | To comparisons of MRDM algorithms Like MR-RADIX vs. Patricia Mine. | Radix Tree Structure | MR-Radix Algorithm is based on the Patricia mine algorithm which can extract multi relational association rules from a large relational database | Tree Structure MR-Radix and Patricia Mine | Relational Database (SIVAT) | Medical data base HC- (Cancer Hospital database responsible for hospital data | JAVA(J2SE6.0)Development Kit from the Net Bean6.7.1 Development Environment and MySql |
| Et.al Arno,J.Knobbe " Involving aggregate functions in Multi Relational Search | When the data is scattered over the many tables which will causes many problems in the data mining .To avoid the above problem , they have used some of the aggregate functions | Generalized Selection Graph and Graph Refinement | Select Sub-Graph Algorithm as well as They have used some of the aggregate function in the database | - | Mutagenesis and Financial database are used | -- | SQL |
| Et.al.Roonie Bathron " Using UML Extensions for specifying domain Knowledge for Data Mining " | They have proposed that how to mine the structured objects by using object oriented terminology. | They have used for the class diagram | An object oriented methodology | UML integration and interaction diagram | - | - | - |
| et.al Biological Application of Multi Relational data Mining | Probabilistic language model to simultaneously predict perimeters and terminators | Hidden Markov and Probabilistic relational model (PRMs) | GOLEM program was used to model the structure activity relationships of trimethoprim analogues binding to dihydrafolate reeducate. RELAGGS was used to convert a multi relational database containing information about genes, gene expressions, proteins and protein-protein interactions into a single table | - | yes | Biological Database | ILP |
| et.al Kavuruchu/Knowledge based System-concept Discovery on relational database: New Techniques for search space pruning and rule quality improvement. | It provides the better rules (Higher accuracy and coverage) can be discovered by using aggregate predicates in the background knowledge. | This paper implements based on the flowcharts--> CRIS Algorithm | CRIS Algorithm | - | Bench Mark Data set(Mesh Mutagenesis, Same gen etc) | - | The SQL queries are used for obtaining the values of support and confidence. |

## 10. COMPARISON BETWEEN EXISTING METHOD (ILP, GM, SSDM) VS MRDM

This section provides the comparison of the above mentioned technique and which is described in the form of table. During comparison so many attribute to be considered these are as below:

1. Attributes:
2. Numeric Values :
3. Intentional Data :
4. Graph / Tree
5. Order in Structural Part
6. Structured terms:

| Property | ILP(Inductive Logic Program) | GM(Graph Mining) | SSDM(Semi-Structured Data Mining) | Multi-Relational Data Mining |
|---|---|---|---|---|
| Attributes | Possible | Not Possible | Yes | Possible |
| Numeric value | Possible | Not Possible | Not Possible | Possible |
| Intentional data | Possible | Not Possible | Not Possible | Possible |
| Graph/tree | Possible for Graph Representation | Possible for Graph Representation | Possible for Tree Representation | Possible for Graph Representation |
| Order in structural parts | Not Possible | Not Possible | Possible | Not Possible |
| Structured terms | Possible | Not Possible | Not Possible | Possible |

**Table 3 – Comparison of (Existing Method) ILP, GM, SSDM Vs Proposed Method (MRDM)**

In the above comparison of the ILP, GM, SSDM and MRDM the credit goes to MRDM .Multi Relational data mining supports all the properties which are mentioned above. As comparisons of ILP, GM, SSDM and MRDM, The favor goes to MRDM.

## 11. CONCLUSION

This paper provides an overview of MRDM approaches to the real world applications and classification methods across multiple database relations including ILP based, Relational database based, Decision tree induction. In this comparative study we conclude that the efficiency and efficiency of the multi relational data mining algorithms for use of relational databases in terms of **execution time and utilized memory.** The above approaches has analyzed with the comparison of traditional data mining algorithm and multi-relational data mining algorithm**.** In fact a better efficiency as it avoids costly multiple joining operations. This paper presents the Multi relational data mining approaches and the methods across multiple database relations including ILP based, Relational database based, Emerging Pattern based, Associative based approaches. Multi-relational data mining deals with knowledge discovery from relational databases consisting of multiple tables. With the development of data mining techniques, multi relational data mining has become a new research area. The researcher can focuses some of the feature direction of the above mentioned applications. MRDM not only deals with the structured and propositional but also discovered the patterns through ILP system.

## 12. ACKNOWLEDEGMENTS

## 14. AUTHOR'S PROFILE

**Mr.Neelamadhab Padhy** is working as an Assistant Professor in the Department of Information and Technology at Gandhi Institute of engineering and Technology (GIET) , India. He has done a post- graduate from Berhampur University, Berhampur, India. He is a life fellow member of Indian Society for Technical Education (ISTE). He is presently pursuing the doctoral degree in the field of Data Mining. He has total teaching experience of 10 years He has a total of 5 Research papers published in National / International Journals into his credit. Presently he has also published 2 Books one is for Programming in C and other is Object Oriented using C++. He has received his MTech (computer science) from Berhampur University Berhampur, 2009.His main research interests are Data warehousing and Mining, Distributed Database System. He is a resource scholar of CMJ University, Meghalaya (Shilong).

**Mrs.Rasmita Panigrahi** is a currently working as a lecturer in the department of information and technology ,Gandhi Institute of Engineering and Technology .She received her MCA from BPUT,(Biju Patanaik University of Technology University ,Rourkela 2010 and she has completed her MTech(Computer Science) in Berhampur University ,Berhampur with a distinction .She has published 4 journals. Her main research interests are Data Warehousing and Mining, Distributed Database System, Designing and Algorithm.